\input{aipcheck}

\documentclass[
    ,final            
  ]
  {aipproc}

\layoutstyle{8x11double}

\begin{document}

\title{Quantum Gravitational Effects and Grand Unification}

\classification{12.10.Kt 04.60.-m 12.10.Dm}
\keywords      {grand unification, supersymmetry, quantum gravity}

\author{Xavier Calmet}{
  address={Catholic University of Louvain,
Center for Particle Physics and Phenomenology,
2, Chemin du Cyclotron,
B-1348 Louvain-la-Neuve, Belgium}
}

\author{Stephen~D.~H.~Hsu}{
  address={Institute of Theoretical Science, University of Oregon,
Eugene, OR 97403, USA}
}

\author{David Reeb}{
  address={Institute of Theoretical Science, University of Oregon,
Eugene, OR 97403, USA} 
}

\begin{abstract}
In grand unified theories with large numbers of fields, renormalization effects significantly modify the scale at which quantum gravity becomes strong. This in turn can modify the boundary conditions for coupling constant unification, if higher dimensional operators induced by gravity are taken into consideration. We show that the generic size of, and the uncertainty in, these effects from gravity can be larger than the two-loop corrections typically considered in renormalization group analyses of unification. In some cases, gravitational effects of modest size can render unification impossible.  
\end{abstract}

\maketitle


LEP data hint towards a unification of the coupling constants of the standard model, or possibly of its supersymmetric version, at a large energy scale of order $10^{16}\,{\rm GeV}$ \cite{Amaldi:1991cn}. However, this scale is uncomfortably close to the Planck scale -- the energy at which quantum gravitational effects become strong. Such effects can alter the boundary conditions on coupling constant unification at the grand unified scale \cite{Hill:1983xh}, and, since their precise size is only determined by Planck scale physics, introduce uncertainties in predictions of grand unification.

We identify an additional uncertainty, arising from the renormalization of the quantum gravity scale itself. We find that the Planck scale is reduced significantly in models with large numbers of particles (e.g., of order  $10^3$ species, common in many grand unified models, and often mostly invisible at low energies). This in turn leads to additional uncertainties in the low energy coupling values associated with unification (see Fig.~1); these uncertainties are generically as large as the two-loop corrections to the renormalization group equations that have become part of the standard analysis of grand unification. Our results suggest that low energy results alone cannot, with any high degree of confidence, either suggest or rule out grand unification.

The strength of the gravitational interaction is modified, i.e.~renormalized, by matter field fluctuations \cite{Larsen:1995ax,Calmet:2008tn}. One finds that the effective Planck mass depends on the energy scale $\mu$ as
$M( \mu )^2 =
M(0)^2 - \frac{\mu^2}{12 \pi} \left(N_0+N_{1/2}-4 N_1\right)
$,
where $N_0$, $N_{1/2}$ and $N_1$ are the numbers of real spin zero scalars, Weyl spinors and spin one gauge bosons coupled to gravity. $M(0)=M_{{\rm Pl}}$ is the Planck mass at low energies -- i.e., it is directly related to Newton's constant $G = M (0)^{-2}$ in natural units $\hbar = c = 1$. Related calculations performed in string theory, which presumably take into account quantum gravity effects, lead to the same behavior for the running of the Planck mass \cite{Kiritsis:1994ta}.

If the strength of gravitational interactions is scale dependent, the true scale $\mu_*$ at which quantum gravity effects are large is one at which
$M (\mu_*) \sim \mu_*$.
This condition means that fluctuations in spacetime geometry at length scales $\mu_*^{-1}$ will be unsuppressed. It has been shown in \cite{Calmet:2008tn}   that the presence of a large number of fields can dramatically impact the value $\mu_*$. For example, it takes $10^{32}$ scalar fields to render $\mu_* \sim {\rm TeV}$, thereby removing the hierarchy between weak and gravitational scales. In many grand unified models, which we study here, the large number of fields can cause the true scale $\mu_*$ of quantum gravity to be significantly lower than the naive value $M_{{\rm Pl}} \sim 10^{19}\,{\rm~GeV}$. In fact, from the above equations,
\begin{equation}
\mu_* = M_{{\rm Pl}}/\eta~,
\end{equation}
where, for a theory with $N \equiv N_0+N_{1/2}-4 N_1$,
\begin{equation}
\eta=\sqrt{1+\frac{N}{12\pi}}~.
\end{equation}

We will exhibit examples of grand unified theories with $N \sim {\cal O}(10^3)$, so that the scale of quantum gravity is up to an order of magnitude below the naive Planck scale. In such models, corrections to the unification conditions from quantum gravity are much larger than previously considered \cite{Hill:1983xh}. In this paper, we show that the generic size of these effects can be larger than the two-loop corrections usually considered in RG analyses of unification, and that in some cases even modestly sized gravitational effects can render unification impossible.  Such large uncertainties might impact whether one considers apparent
unification of couplings to be strong evidence for grand unification
or supersymmetry.

The breaking of a grand unified gauge group down to the standard model group SU(3)$\times$SU(2)$\times$U(1) via Higgs mechanism typically involves several scalar multiplets, which can be in large representations. Furthermore, the total number of these scalar degrees of freedom in the form of Higgs bosons is typically much larger than the number of gauge bosons, so $N=N_0+N_{1/2}-4 N_1$ can be large. In this paper, we mainly consider supersymmetric grand unified theories, since they naively lead to better unification results compared to non-supersymmetric models \cite{Amaldi:1991cn}.

For example, SUSY-SU(5) with three families already has $N=165$, i.e.~$\eta=2.3$. In SUSY-SO(10) models, the numbers are larger: the minimal supersymmetric SO(10) model  uses ${\bf 126}$, ${\bf \overline{126}}$, ${\bf 210}$ and ${\bf 10}$ Higgs multiplets, yielding $N=1425$ or $\eta=6.2$. Some models use even more multiplets, others use fewer and smaller ones, although the model with the smallest representations ${\bf 10}$, ${\bf 16}$, ${\bf \overline{16}}$ and ${\bf 45}$  -- yielding $N=270$ and $\eta=2.9$ -- leads to R-parity violation and other problems. We thus have $\eta \sim 5$ for most reasonable SUSY-SO(10) models. Other unification groups considered in the literature include ${\rm E8}\times{\rm E8}$, which is motivated by string theory and requires both ${\bf 248}$ and ${\bf 3875}$ Higgs multiplets, clearly yielding even bigger renormalization effects on $M_{{\rm Pl}}$.

Quantum gravity effects have been shown to affect the unification of gauge couplings \cite{Hill:1983xh}. The lowest order effective operators induced by a quantum theory of gravity are of dimension five, such as \cite{Hill:1983xh}
\begin{eqnarray}
\label{dim5} 
\frac{c}{\hat{\mu}_*} {\rm Tr}\left(G_{\mu\nu} G^{\mu\nu} H\right)~,
\end{eqnarray}
where $G_{\mu\nu}$ is the grand unified theory field strength and $H$ is a scalar multiplet. This operator is expected to be induced by strong non-perturbative effects at the scale of quantum gravity, so has coefficient $c \sim {\cal O}(1)$ and is suppressed by the reduced true Planck scale $\hat{\mu}_*=\mu_*/\sqrt{8\pi}=\hat{M}_{{\rm Pl}}/\eta$ with $\hat{M}_{{\rm Pl}}=2.43\times 10^{18}\,{\rm GeV}$. Note, there is some ambiguity as to whether the Planck scale $\mu_*$ or the reduced Planck scale $\hat{\mu}_*$ should be used \cite{Hill:1983xh}. Our main point here is the gravitational enhancement $\eta$ of this operator due to renormalization of the quantum gravity scale, which has not been taken into consideration previously.

To be as concrete and unambiguous as possible, we will first examine these gravitational effects in the example of SUSY-SU(5). Operators similar to (\ref{dim5}) are present in all grand unified theory models and an equivalent analysis applies. The following analysis can be carried over verbatim to SO(10) models \cite{Calmet:2008df}.

In SU(5) the multiplet $H$ in the adjoint represenation acquires, upon symmetry breaking at the unification scale $M_X$, a vacuum expectation value $\left\langle H \right\rangle = M_X \left(2,2,2,-3,-3\right)/\sqrt{50\pi\alpha_G}$, where $\alpha_G$ is the value of the SU(5) gauge coupling at $M_X$. Inserted into the operator (\ref{dim5}), this modifies the gauge kinetic terms of SU(3)$\times$SU(2)$\times$U(1) below the scale $M_X$ to
\begin{eqnarray}
\label{gaugekineticterm}
-\frac{1}{4} \left(1+\epsilon_1\right)F_{\mu\nu} F^{\mu\nu}_{{\rm U}(1)}
-\frac{1}{2}\left(1+\epsilon_2\right){\rm Tr}\left(F_{\mu\nu} F^{\mu\nu}_{{\rm SU}(2)}\right)\\  \nonumber
 -\frac{1}{2}\left(1+\epsilon_3\right){\rm Tr}\left(F_{\mu\nu} F^{\mu\nu}_{{\rm SU}(3)}\right)
\end{eqnarray}
with
$
\epsilon_1=\epsilon_2/3=-\epsilon_3/2=\frac{\sqrt{2}}{5\sqrt{\pi}}\frac{c\eta}{\sqrt{\alpha_G}}\frac{M_X}{\hat{M}_{{\rm Pl}}}$.

After a finite field redefinition $A_{\mu}^{i} \to \left(1+\epsilon_i\right)^{1/2} A_{\mu}^{i}$ the kinetic terms have familiar form, and it is then the corresponding redefined coupling constants $g_i \to \left(1+\epsilon_i\right)^{-1/2} g_i$ that are observed at low energies and that obey the usual RG equations below $M_X$, whereas it is the \emph{original} coupling constants that need to meet at $M_X$ in order for unification to happen. In terms of the observable rescaled couplings, the unification condition therefore reads:
\begin{eqnarray}
\label{boundarycondition}
\alpha_G  &=& \left(1+\epsilon_1\right) \alpha_1(M_X)=\left(1+\epsilon_2\right) \alpha_2(M_X) \\ \nonumber
 &=& \left(1+\epsilon_3\right) \alpha_3(M_X)~.
\end{eqnarray}

Numerical unification results using this boundary condition are shown in Fig.~1. Leaving the parameters $\alpha_3(M_Z)$ and $M_{{\rm SUSY}}$ open in some range in order to compare the size of the corrections from the new boundary condition to experimental uncertainties, we evolved the gauge couplings under two-loop RG equations of the SM/MSSM with SUSY breaking scale $M_{{\rm SUSY}}$, taking as fixed $\alpha_1(M_Z)=0.016887$, $\alpha_2(M_Z)=0.03322$. Then, testing each pair ($\alpha_3(M_Z)$, $M_{{\rm SUSY}}$) in the wide range of parameters of Fig.~1 for unification according to (\ref{boundarycondition}), it turns out that for every pair perfect unification happens for exactly one value of the coefficient $c$ of (\ref{dim5}).

\begin{figure}[htb]
\resizebox{.43\textwidth}{!}
{\includegraphics[width=1\linewidth]{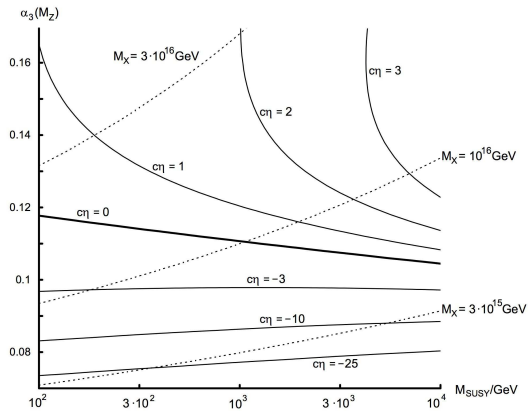}}
\caption{For $\eta$ fixed by the particle content of the theory, solid lines are contours of constant $c$ such that, under the presence of the gravitationally induced and enhanced operator (\ref{dim5}), SUSY-SU(5) perfectly unifies at two loops for given values of the initial strong coupling constant $\alpha_3(M_Z)$ and SUSY breaking scale $M_{{\rm SUSY}}$. Over the whole range, unification happens for some value of the coefficient $c$, with unification scale and unified coupling between $M_X=9.3\times 10^{14}\,{\rm GeV}$, $\alpha_G=0.033$ (lower right corner) and $M_X=5.5\times 10^{16}\,{\rm GeV}$, $\alpha_G=0.045$ (upper left).}
\end{figure}

Our results show that, e.g., in a theory with $\eta \sim 5$, unification depends quite sensitively on the size of the gravitational operator: reasonable values of the coefficient $c \sim {\cal O}(1)$ can give unification for quite a large range of low-energy couplings $\alpha_i(M_Z)$ and parameters $M_{{\rm SUSY}}$, so unification does not seem to be a very special feature. On the other hand, even a slight change to the value of $c$ requires quite large adjustments in initial conditions $\alpha_i(M_Z)$ for unification to still happen. This is very unsatisfying since the value of $c$ is determined only by some deeper theory of quantum gravity above the scale $M_X$, i.e.,  grand unification cannot be predicted or claimed based on low-energy observations alone, and therefore loses most of its beauty. More severely yet, the effects of the gravitational operator can be so large that, if quantum gravity determines the sign of this operator to be positive with $c>4/\eta$ (which is quite natural for theories with large particle content), then unification cannot happen for any experimentally allowed parameters of the SM/MSSM model, see Fig.~1.

Improving the precision of theoretical predictions and experimental values seems unnecessary and meaningless: e.g., for the parameter values $\alpha_3(M_Z)=0.108$, $M_{{\rm SUSY}}=10^3\,{\rm GeV}$, $M_X=10^{16}\,{\rm GeV}$ and $\alpha_G=0.0389$ favored by Amaldi {\it et al.}~\cite{Amaldi:1991cn} to yield good unification, table I compares the shifts $\alpha_i^{2}(M_X)-\alpha_i^{1}(M_X)$ in theoretical predictions due to inclusion of two-loop running to the splittings $\alpha_G-\alpha_G/(1+\epsilon_i)$ required by the the boundary condition (\ref{boundarycondition}). These splittings are shown for $\eta \sim 5$ and $c=-1$, but would be larger or smaller proportional to $c\eta$. The table shows that the generic size of, and uncertainty in, the effects from gravity is larger than the two-loop corrections. Thus, two-loop computations do actually not improve evidence for unification.

\begin{table}[!thb]
\resizebox{0.9\columnwidth}{!}
{\begin{tabular}{c|c|c|c}
\hline
$i$ & 1 & 2 & 3 \\
\hline
$\alpha_i^{1}(M_X)$ & 0.03815 & 0.03767 & 0.03814 \\
$\alpha_i^{2}(M_X)$ & 0.03897 & 0.03899 & 0.03868 \\
$\delta\alpha_i=\alpha_i^{2}-\alpha_i^{1}$ & $8.2\times 10^{-4}$ & $13.2\times 10^{-4}$ & $5.4\times 10^{-4}$ \\
$\delta\alpha_i/\alpha_i^{1}$ & $+2.1\%$ & $+3.5\%$ & $+1.4\%$ \\
\hline
$\epsilon_i(c\eta=-5)$ & $-0.0167$ & $-0.0503$ & $+0.0335$ \\
$\alpha_G(M_X)$ & 0.0389 & 0.0389 & 0.0389 \\
$\alpha_{Gi}=\alpha_G/(1+\epsilon_i)$ & 0.0396 & 0.0410 & 0.0376 \\
$\delta_i=\alpha_G-\alpha_{Gi}$ & $-6.6\times 10^{-4}$ & $-20.6\times 10^{-4}$ & $12.6\times 10^{-4}$ \\
$\delta_i/\alpha_G$ & $-1.7\%$ & $-5.3\%$ & $+3.2\%$ \\
\hline
\end{tabular}}
\caption{The upper half of the table shows shifts in the predictions for the values of the coupling constants at $M_X=10^{16}\,{\rm GeV}$ due to inclusion of two-loop running. These shifts are comparable in size or even smaller than the necessary splittings between the $\alpha_{Gi}$ due to (\ref{boundarycondition}) in the case $\eta=5$, $c=-1$ (lower half).}
\end{table}

Similarly, the uncertainty in the value of the coefficient $c$ is far greater than experimental uncertainties in measurements of SM/MSSM parameters. For example, the parameter range $\alpha_3(M_Z)=0.108 \pm 0.005$, $M_{{\rm SUSY}}=10^{3 \pm 1}\,{\rm GeV}$ quoted in \cite{Amaldi:1991cn} is covered by varying the coefficient $c$ in the small range $-2/\eta < c < 2/\eta$, see Fig.~1. In particular, previous attempts to pin down $\alpha_3(M_Z)$ or $\sin^2 \theta_W$ by demanding gauge coupling unification seem invalid without further knowledge about $c$. Also, claiming that SUSY unification is favored by, e.g., LEP data seems far-fetched. Without actually observing proton decay it is hard to claim convincing evidence for unification of the gauge interactions of the standard model at some higher scale. Finally, as can be seen from Fig.~1, the unification scale that would be compatible with current experimental values of $\alpha_3(M_Z)$ is of the order of $M_X \sim 10^{16}\,{\rm GeV}$, which, depending on the specific model under consideration, might be uncomfortably low with respect to proton decay. Phrased another way,  given the current measurements of $\alpha_i(M_Z)$, the operator (\ref{dim5}) cannot be used to shift the unification scale $M_X$ to values much above $10^{16}\,{\rm GeV}$ (this possibility was discussed in past analyses \cite{Hill:1983xh}).

Many predictions of grand unified theories are subject to uncertainties due to quantum gravitational corrections. We have shown that these uncertainties are significantly enhanced in models with large particle content (e.g., of order $10^3$ matter fields), including common variants of SU(5), SO(10) and ${\rm E8}\times{\rm E8}$ unification. Since the quantum gravitational corrections and, potentially, most of the large number of matter fields appear only at very high energies, it seems that low energy physics alone cannot, with a high degree of confidence, either suggest or rule out grand unification. Model builders should perhaps favor smaller matter sectors in order to minimize these corrections and obtain calculable, predictive results.


\begin{theacknowledgments}
  This work is supported in part by the
Department of Energy under DE-FG02-96ER40969. X.C. is a "Charg\'e de recherches du F.R.S.-FNRS."
\end{theacknowledgments}

\end{document}